\documentclass[12pt]{svjour3}                     
\usepackage{graphicx}
\usepackage{amsmath,amsfonts,amssymb}
\usepackage{bm}
\usepackage{mathrsfs}
\journalname{General Relativity and Gravitation}

\newcommand{\beq}{\begin{equation}}
\newcommand{\eeq}{\end{equation}}
\newcommand{\beqa}{\begin{eqnarray}}
\newcommand{\eeqa}{\end{eqnarray}}
\def\nn{\nonumber\\}
\def\eq#1{(\ref{#1})}

\def\cd#1{\ensuremath{\nabla_{#1}}}          
\def\pd#1{\ensuremath{\partial_{#1}}}        

\newcommand{\lag}{\ensuremath{\mathscr{L}}}        

\def\dlag#1{\frac{\partial\lag}{\partial\ \! #1}}

\def\st{spacetime}

\def\mch{\scriptscriptstyle}

\def\text#1{{\rm #1}}

\def\lab#1{\label{#1}}

\def\altura{\rule{0pt}{16pt}}
\def\E{\mathcal{E}}
\def\R{\mathbb{R}}

\begin{document}

\title{Higher-dimensional perfect fluids and empty singular boundaries}


\titlerunning{Higher-dimensional perfect fluids and empty singular boundaries}        

\author{Ricardo\ E.  Gamboa Sarav\'{\i}      
}

\authorrunning{R. E.  Gamboa Sarav\'{\i}} 

\institute{R. E. Gamboa Sarav\'{\i} \at
              Departamento de F\'{\i}sica, Facultad de Ciencias
Exactas, Universidad Nacional de La Plata, and
 IFLP, CONICET. C.C. 67, (1900) La Plata, Argentina.  \\
                            \email{quique@fisica.unlp.edu.ar}           
}

\date{Received: date / Accepted: date}

\maketitle

\begin{abstract}
In  order to find out whether empty singular boundaries can arise in higher dimensional Gravity,
we study the  solution of Einstein's equations  consisting in a ($N+2$)-dimensional  static and hyperplane symmetric
perfect fluid satisfying the equation of state $\rho=\eta\, p$, being $\rho$ an arbitrary constant and $N\geq2$.

We show that this
spacetime has
some weird  properties. In particular, in the case $\eta>-1$, it has an empty (without matter) repulsive singular boundary.

We also study the behavior of geodesics and  the Cauchy problem for the propagation of massless scalar field in this spacetime.
For $\eta>1$,  we find that only  vertical null geodesics touch the boundary and
bounce, and  all of them start and finish at $z=\infty$; whereas non-vertical
null as well as all time-like ones are bounded between two planes determined by initial conditions.
We obtain that the Cauchy problem for the propagation of a massless scalar field is well-posed and waves are completely reflected at the singularity, if we only demand the waves to have finite energy, although no boundary condition is required.

\end{abstract}

\section{Introduction}

We have recently pointed out that solutions of the Einstein field equations
with empty singular boundaries
can exist in $ 4 $ dimensions \cite {q1,q2,q3,q4}. These singularities are not the sources of the fields, but rather, they arise due to the attraction of distant matter.

In such spacetimes, at the boundary, some components of the Weyl  tensor diverge while the energy-momentum tensor vanishes.
It turns out that only orthogonal null geodesics just touch the singularity and bounce, whereas non-orthogonal null ones as well as all time-like ones bounce before getting to it. Furthermore, we  have shown that the Cauchy problem  for wave propagation  is well-posed, if we only demand the waves to have finite energy, although no boundary condition is required \cite{GST}. And  waves are completely reflected at the singularity. Due to these properties we also call this kind of singularities {\em white walls}\footnote{A reviewer suggested for them the name {\em singular mirrors} instead {\em white walls}.}.

Needless to say, the possibility that space-time may have more than four dimensions is now a
standard assumption in High Energy Physics. Indeed,  most theories attempting to unify all the fundamental interactions  require a higher dimensional spacetime.

However, it is well known that Einstein's Gravity gets weaker as  the number of \st\  dimensions increases. For instance,  stable bound Keplerian orbits do not exist in \st s of more than $4$ dimensions  \cite{Tang}, or also, the degree of compactification of spherical stars diminishes as the dimensionality of \st\  raises \cite{Cruz}.

Thus, it would be worthwhile   finding out whether  empty singular boundaries can arise in higher dimensional Gravity,  and the aim of this paper is to explore this issue.

In this work, we  show that such  singularities may occur in \st s of $N+2$ dimensions ($N\ge2$) by finding an explicit solution.

The solution to be described is a ($N+2$)-dimensional static and hyperplane symmetric \st\ filled with a perfect fluid satisfying the equation of state  $\rho=\eta\, p$, being $\eta$ an arbitrary constant.

In section \ref{sol}, we find the solution and describe its main features depending on the value of $\eta$ and $N$. In particular we show that, for any $N\ge2$, this spacetime has an empty singular boundary if $\eta>-1$. We also discuss the vacuum and dust limits.

In section \ref{geo},   we study the properties of null and time-like geodesics for the most physically interesting case $\eta>1$.

In section \ref{Cauchy},   we discuss well-poseness of Cauchy problem for the propagation of a massless field in this case.

Throughout this paper, we adopt the convention in which the \st\
metric has signature $(-\ +\ \dots +\ +)$, the  system of units in which
the speed of light $c=1$, and $G$ is the $(N+2)$-dimensional  Newton's gravitational
constant.

\section{The solution}\lab{sol}

Let us  consider a solution of Einstein's field
equations corresponding to a static and hyperplane symmetric
distribution of matter. That is, they must be
invariant under $n$ translations and under rotations in $N(N-1)/2$ planes. The matter we will consider is a perfect fluid satisfying the equation of state
\beq\lab{eos}
\rho=\eta\, p\,,
\eeq
where $\eta$ is an arbitrary  constant. Then the components of the stress-energy tensor read
\beq T_{ab}= p\,(\eta+1)\,u_au_b+p\, g_{ab}\, ,\eeq
where $u^a$ is
the velocity of fluid elements.

Due to the required symmetry, by solving Killing's equations it can be readily seen that coordinates can chosen  such that the line element can be written as \footnote{See appendix \ref{A}.}
 \beq \lab {met} ds^2= - \mathcal{G}(z)^2\,dt^2+ e^{2V(z)}\left((dx^{\mch
1})^2+\dots+(dx^{\mch{N}})^2
\right)+dz^2\,.\eeq%

The non identically vanishing components of the Einstein tensor
are
\beqa \lab {gtt} G_{tt}=-N\,\mathcal{G}^2 \left( V''+\,\frac{N+1}{2}\ V'^2\right),\\
\lab {gii} G_{11}=\dots=G_{NN}= e^{2V} \left(\frac{\mathcal{G}''}{\mathcal{G}}+(N-1)\frac{\mathcal{G}'}{\mathcal{G}}\,V'+(N-1)V''+\frac{N(N-1)}{2}V'^2\right)\,,\\
\lab {gzz} G_{zz}= N\ V' \left( \frac{\mathcal{G}'}{\mathcal{G}}+
\frac{N-1}{2} V' \right)\ ,
 \eeqa where a prime $(')$ denotes differentiation with respect to $z$.%

On the other hand, since the fluid must be static,
$u_a=(-\mathcal{G},0,\dots,0)$,  so \beq T_{ab}=
\text{diag}\left(\rho\, \mathcal{G}^{2},p\, e^{2V},\dots,p\,
e^{2V},p\right)\,, \eeq where  $\rho$ and $p$  depend only on  $z$. Note that due to the imposed symmetry, rotation, expansion and shear of the fluid  velocity vanish, while the acceleration vector field is $\dot{u}_a=(0,\dots,0,\mathcal{G}'/\mathcal{G})$.

Thus, Einstein's equations, i.e., $G_{ab}=4 \pi G \frac{N}{N-1}\ T_{ab}$ \footnote {We have set the normalization constant such that, in the Newtonian limit, they lead to $\nabla^2 \Phi=-\frac12 \nabla^2(1+g_{tt})= 4 \pi G \rho$.}, become
\beqa \lab {gtt1} V''+\frac{N+1}{2}\, V'^2= - \frac{4 \pi\, G}{N-1}\ \rho\,, \\%
\lab {gii1}
\frac{\mathcal{G}''}{\mathcal{G}}+(N-1)\frac{\mathcal{G}'} {\mathcal{G}}\,V'+(N-1)V''+\frac{N(N-1)}{2}V'^2
=\frac{4 \pi N\,G }{N-1}\ {p}\, ,\\
\lab {gzz1}   \frac{\mathcal{G}'}{\mathcal{G}}\,V' +
\frac{(N-1)}{2} V'^2  = \frac{4 \pi\, G}{N-1}\ {p}\,.
 \eeqa%

Moreover, $\cd a T^{ab}=0$ yields
\beq \lab{ppr} p' = -(\rho+p)\,\frac{\mathcal{G}'}{\mathcal{G}}\,.\eeq Of course, due to Bianchi's identities equations
(\ref{gtt1})--(\ref{ppr}) are not
independent.

Some comments are in order. First, it is interesting to see that the $(N+2)$-dimensional Raychaudhuri equation
\beq \dot{\theta}+\frac 1{N+1}\ \theta^2 -\cd a\dot{u}^a + \sigma^{ab}\sigma_{ab}-\omega^{ab}\omega_{ab}+4 \pi G\left(\rho+\frac{N+1}{N-1}\,p\right)=0\ ,\eeq
becomes in this case
\beq \cd a\dot{u}^a =\frac{\mathcal{G}''}{\mathcal{G}}+N\,\frac{\mathcal{G}'} {\mathcal{G}}\,V'=4 \pi G\left(\rho+\frac{N+1}{N-1}\,p\right)\ ,\eeq
which is equivalent to the linear combination $-(N-1)$\eq{gtt1}+\eq{gii1}+$N$\eq{gzz1} of  equations \eq{gtt1}-\eq{gzz1}.

Second, it readily follows from (\ref{gtt1})-(\ref{ppr}) that, for arbitrary constants $A$, $B$, $\lambda\neq0$ and $z_0$, field equations are invariant under the transformation $z\rightarrow \lambda z+z_0$ (i.e., $z$-translations, $z$-dilatations and mirror reflections across any ``plane''), $\mathcal{G}\to A\,\mathcal{G}$, $V\to V+B$, $\rho\to\lambda^2\rho$ and $p\to \lambda^2p$. Thus, if
$\{\mathcal{G}(z),V(z),\rho(z),p(z)\}$ is a solution of (\ref{gtt1})-(\ref{gzz1})  $\{A\,\mathcal{G}(\lambda z+z_0),V(\lambda z+z_0)+B,\lambda^2\rho(\lambda z+z_0),\lambda^2p(\lambda z+z_0)\}$ is another one.

Third, regarding the mirror symmetry, independently of the equation of state and dimension of \st , a solution cannot have a ``plane" of symmetry in a region where $\rho(z)\neq0$
 and $p(z)>0$. In order to see this, let us assume that $z=z_s$ is that ``plane", then it
must  hold that $ \mathcal{G}'= V' =p' =\rho'=0$ at $z_s$, and so from
\eq {gzz1} we get that also $p(z_s)=0$. Now, by differentiating \eq{ppr} and using
\eq{gtt1} and \eq{gii1}, we obtain $p''(z_s)=-4\pi G\,\rho^2<0$.
By using the equation of state we can go further. Indeed, it follows from \eq{eos},\eq{gtt1}-\eq{gzz1} that the four functions $ \mathcal{G},\, V,\, p$ and $\rho$, and all their derivatives of any order must vanish
at $z_s$.

We now proceed to find the solution. By using the  equation  of state \eq{eos}, we readily find from \eq {ppr} that
\beq\lab{presion}{p} = p_0\,
\mathcal{G}^{\,-(\eta+1)},  \eeq where $p_0$ is an arbitrary  constant.
Now, this last expression and the change of variables \beq\lab{xi} \xi = \int_{z_0}^{z}
\mathcal{G}(z')^{\,-\frac{\eta+1}{2}} dz', \eeq
brings \eq{gtt1} and \eq{gzz1} to
\beq  -\frac{\eta+1}{2}\ \frac{\mathcal{G}_\xi}{\mathcal{G}}\,V_\xi +V_{\xi\xi}+\frac{N+1}{2}\, V_\xi^2= - \frac{4 \pi\, G}{N-1}\, \eta\,p_0 \, \eeq and
\beq \lab{gzz2} \frac{\mathcal{G}_\xi}{\mathcal{G}}\,V_\xi +\frac{N-1}{2}\, V_\xi^2=  \frac{4 \pi\, G}{N-1}\, p_0 \,. \eeq
Hence, we find that $V$ satisfies
\beq \lab {gtt4}   V_{\xi\xi} + \frac{\eta(N-1)+3N+1}{4}\  V_\xi^2 +\frac{2 \pi\, G}{N-1} \,(\eta-1)\ p_0 =0,
\eeq%
and, by setting
\beq \lab{alfa}{\alpha}=  \frac{(\eta-1)\bigl(\eta(N-1)+3N+1\bigr)}{2(N-1)}\, \pi\, G\, p_0\ ,\eeq
we can write its general solution  as
 \beq \lab{V1} V(\xi)=\ln\left(
{C_1\,\sin(\sqrt{\alpha}\,\xi+C_2)}\right)^{\frac{4}{\eta(N-1)+3N+1}}\,,\eeq
where $C_1$ and $C_2$ are arbitrary constants.

Now,  we can write  \eq{gzz2} as
\beq \frac{d}{d\xi}\ln\left(\mathcal{G} \right)=\frac{4 \pi\, G}{N-1}\, \frac{C}{V_\xi}-\frac{N-1}{2}\, V_\xi  \,, \eeq
which, by using \eq{V1}, can be readily integrated, giving  the general solution for $\mathcal{G}(\xi)$%
\beq \label{G} \mathcal{G}(\xi)=
\,\left({C_3\,\sin(\sqrt{\alpha}\,\xi+C_2)}\right)^{-\frac{2(N-1)}{\eta(N-1)+3N+1}}{\left(\cos(\sqrt{\alpha}\,\xi+C_2)\right)}^{-\frac{2}{\eta-1}},\eeq
where $C_3$ is a new arbitrary constant.

Now, we define a new positive coordinate $u$ such that\beq \lab{u}u=
\frac{\sin^2(\sqrt{\alpha}\,\xi+C_2)}{\beta}, \eeq and, without losing generality, we set $C_1=C_3={1}/{\sqrt{\beta}}$, being $\beta$ is an  arbitrary  constant. And, taking into account that
\beq\frac{du}{dz}=\frac{du}{d\xi}\ \frac{d\xi}{dz}=\sqrt{\frac{4\alpha}{\beta}}\ u^{\frac{1}{2}}\,(1-\beta u)^{\frac{1}{2}}\ \mathcal{G}^{\,-\frac{\eta+1}{2}}\,,\eeq
we can write the the metric \eq{met} in terms of $u$ as
\beqa \lab{met3}  ds^2=-
u^{-\frac{2(N-1)}{\eta(N-1)+3N+1}}\,\left(1-\beta
u\right)^{-\frac{2}{\eta-1}}\, dt^2 + u^{\frac{4}{\eta(N-1)+3N+1}}\left((dx^{\mch
1})^2+\dots+(dx^{\mch{N}})^2
\right)\nn\ +\,\frac{\beta}{4\alpha}\ \ u^{-\frac{2\eta(N-1)+4N}{\eta(N-1)+3N+1}}\ \  \left(1-\beta
u\right)^{-\frac{2 \eta}{\eta-1}}\, du^2, \hspace{2cm}\nn
-\infty<t<\infty,\;-\infty<x^{\mch 1}<\infty,\;\dots,\;-\infty<x^{\mch N}<\infty,\;0<u\begin{cases}<1/\beta &\text{if}\ \beta>0\\ <\infty &\text{if}\ \beta<0\end{cases}\,.\hspace{.5cm}
\eeqa
For the case $N=2$, this solution was found in \cite{Bron}, see also \cite{BronKov,Coll,SKMHH}; and a more detailed study of its properties was done in \cite{q2}.
Note that, for a  Lorentz manifold, $\alpha$ and $\beta$ must have the same sign  and  none of them can be zero or infinity (see also \eq{u}), therefore $0<\alpha\beta<\infty$. On the other hand, it follows from \eq{u} that, if $\alpha>0$ the coordinate $u$ is bounded ($0<u<1/\beta$), whereas if $\alpha<0$ it is unbounded ($0<u<\infty$).

Note that, by rescaling the coordinates, the parameter $\alpha$ might be eliminated. However, we prefer to keep it and think of it as a scale with dimensions of $(length)^{-2}$. In this way, the parameter $\beta$ and the coordinate $u$ turn out to be dimensionless. In this sense, for each $N>1$, the solution essentially depends on only two dimensionless parameters $\beta$ and $\eta$.

Even though  in this work we are mainly interested in the case $\eta>1$, it is worth mentioning that \eq{met3} has a much wider range of validity. Indeed,  for each $N>1$, from the mathematical point of view the solution given in \eq{met3} makes sense for arbitrary finite values of the parameters, provided $\alpha\beta>0$, $\eta\neq1$ and  $\eta\neq-\frac{3N+1}{N-1}$. Even more, for these excluded values of $\eta$ the solution can  be found as limit cases of \eq{met3}.
Nevertheless, for the sake of completeness, we shall give here a briefly  description of the main features of the solution for arbitrary  values of $\eta$.

In what follows, let us assume that $0<\alpha\beta<\infty$, $|\eta|<\infty$, $\eta\neq1$ and  $\eta\neq-\frac{3N+1}{N-1}$.
In view of \eq{presion},\eq{alfa} and \eq{met3}, the fluid  density   may be written as
\beq
\lab{ro}\rho(u)=\eta p(u)= \frac{2\,\eta(N-1)\,\alpha}{(\eta-1)\bigl(\eta(N-1)+3N+1\bigr)\pi G}\  u^{\frac{(\eta+1)(N-1)}{\eta(N-1)+3N+1}}\,\left(1-\beta u\right)^{\frac{\eta+1}{\eta-1}}. \eeq
Hence, the sign  $\alpha$ must to be chosen so that ${\alpha\,\eta}{(\eta-1)\bigl(\eta(N-1)+3N+1\bigr)}$  has the same sign as the fluid density, and this choice sets the sign of $\beta$ as well.

Therefore, for  $\rho>0$, the parameter $\alpha$ is positive for the cases $\eta>1$ or $0>\eta>-\frac{3N+1}{N-1}$, but negative for $1>\eta>0$ or $-\frac{3N+1}{N-1}>\eta$. Thus the range of the coordinate $u$ is ($0<u<1/\beta$) for the former cases and ($0<u<\infty$) for the latter ones. On the other hand, it follows from \eq{ro}  that $\rho(u)$ and $p(u)$ are $C^\infty$  functions of $u$ in the interior of the range. The pressure is positive for $\eta>0$ and negative for $0>\eta$, and the dominant energy condition $|p|/\rho\leq1$ holds only for $\eta\geq1$ or $-1\geq\eta$.

For $\eta=-1$, which is a case with $\alpha>0$, it follows from \eq{ro} that $\rho=-p=\frac{(N-1)\alpha}{2(N+1)\pi G}$. Hence, in this case, \eq{met3} turns out to be a vacuum solution with cosmological constant $\Lambda=\frac{2N}{(N+1)}\,\alpha$; this is the higher-dimensional generalization of that found in \cite{NH}.

For $\eta\ne-1$, it follows from \eq{ro} that at the endpoints of the range of $u$, depending on the sign of the exponents  of $u$ and $(1-\beta u)$, $p(u)$ either vanishes or diverges. At $u=0$ the exponent is $\frac{(\eta+1)(N-1)}{\eta(N-1)+3N+1}$, so  it vanishes for the cases $\eta>-1$ or $-\frac{3N+1}{N-1}>\eta$, but diverges for $-1>\eta>-\frac{3N+1}{N-1}$. Thus, this property together with the sign of $\alpha$ define five regions where the solutions with positive $\rho$ have very different features. Let us enumerate these regions in the following way,  (I): $\eta>1$, (II): $1>\eta>0$, (III): $0>\eta>-1$, (IV): $-1>\eta>-\frac{3N+1}{N-1}$, and (V): $-\frac{3N+1}{N-1}>\eta$. In  Table \ref{tabla} we list the main properties of the solution for the different regions.

Similarly it follows from \eq{ro} that, at the other endpoint ($u=1/\beta$ or $u=\infty$, depending on the sign of $\alpha$),  $p(u)$ vanishes for the cases (I), (II) and (IV), but diverges for the  (III) and (V) ones (see Table \ref{tabla}).

Note that the proper  distance coordinate $z$  used in \eq{met}, is implicitly given in terms of $u$ by
\beqa \lab{z} z&=& \pm\int_{u_0}^{u}\sqrt{g_{\mch uu}(u')}\, du' 
=\pm\sqrt{\frac{\beta}{4\,\alpha}}\,\int_{u_0}^{u}\,(u')^{\frac{N+1}{\eta(N-1)+3N+1}-1}\,\left(1-\beta u'\right)^{-\frac{1}{\eta-1}-1}\, du'\nn
&\to& \begin{cases}C +C'\ u^{\frac{N+1}{\eta(N-1)+3N+1}}\ \bigl(1+O(u)\bigr)&\text{as}\,\,\, u\rightarrow0,\\ 
C +C'\left(1-\beta\,
u\right)^{-\frac{1}{\eta-1}}\ \bigl(1+O(1-\beta u)\bigr)& \text{for}\ \beta>0\ \text{as}\,\,\, u\rightarrow1/\beta,
\\
C +C'\ u^{\frac{N+1}{\eta(N-1)+3N+1}-1-\frac{1}{\eta-1}}\ \bigl(1+O(u^{-1})\bigr)&\text{for}\ \beta<0\ \text{as}\,\,\, u\rightarrow\infty,\
\end{cases}\  \eeqa 
where $C$ and $C'$ are given constants for each of the cases.

When $u\to0$ it follows from \eq{z} that, $z\to z_0$ in cases (I), (II), (III) and (IV), and $z\to\pm\infty$ in case (V). At the other endpoint,  $z\to\pm\infty$ in cases (I) and (II), and
$z\to z_1$ in cases (III), (IV) and (V).
Therefore,  spacetime is semi-infinite in the z-direction in cases (I), (II) and (V), while it is bounded between two boundaries in cases (III) and (IV).

\begin{table}\begin{center}
\begin{tabular}{|c|c|c|c|c|c|}   \hline

\rule{0pt}{10pt}Region&I&II& III&IV&V\\
\hline\hline\altura

$\eta$&$\eta>1$&$1>\eta>0$&$0>\eta>-1$& $-1>\eta>-\frac{3N+1}{N-1}$&$-\frac{3N+1}{N-1}>\eta$\\
\hline\altura

$\alpha,\beta$&$+$& $-$&$+$&$+$&$-$\\
\hline\altura

  range of $u$&$(0,1/\beta)$&$(0,\infty)$&$(0,1/\beta)$&$(0,1/\beta)$&$(0,\infty)$\\
\hline\altura

$p$&$+$& $+$&$-$&$-$&$-$\\
\hline\altura

$|p|/\rho$&$<1$&$ >1$&$>1$&$<1$&$<1$\\
\hline\altura

$u\to 0^+$&$p\to0$&$p\to0$&$p\to0$&$p\to -\infty$&$p\to0$\\
\hline\altura

$u\to(1/\beta)^-$ &$p\to0$& --- &$p\to-\infty$&$p\to0$&---\\
\altura

$u\to\infty$ &---& $p\to0$&---&---&$p\to-\infty$\\
\hline\altura

$u\to0$ &$\to z_0$ & $\to z_0$  &$\to z_0$&$\to z_0$&$\to\pm\infty$\\
\hline \altura

$u\to1/\beta$&$\to \pm\infty$ &---&$\to z_1$&$\to z_1$&---\\
\altura
$u\to\infty$&---& $\to \pm\infty$ &---&---&$\to z_1$\\
\hline\altura


$u=0$&singular& singular &singular&singular&regular\\
\hline\altura

$u=1/\beta$ &regular& --- &singular&regular&---\\
\altura

$u=\infty$ &---& regular &---&---&singular\\
\hline

\end{tabular}\end{center}\caption{Some properties of solutions with positive $\rho$ according to the value of $\eta$.\lab{tabla}}
  \end{table}

The solution may  have a \st\, curvature singularity only at the endpoints of the range of $u$, since straightforward computation of the Kretschmann curvature invariant yields%
\beqa\lab{RR} K= R_{abcd}R^{abcd}=\frac{256\ \alpha^2 \ u^{-\frac{4 (N+1)}{\eta(N-1)  +3 N+1}}\  (1-\beta\,u
)^{{2\frac{ \eta+1  }{\eta -1 }}}}{{\beta ^2\,(\eta-1)^2\,  \bigl(\eta (N-1) +3 N+1\bigr)^4}}\, \Bigl(N^2  (N^2-1)\, (\eta -1)^2\nn
+ 2N^2  (N-1)  (\eta^2 -1)\, \bigl((N-1)\, \eta +N+1\bigr)\,\beta \, u +
P(N,\eta)\,\beta^2 u^2\Bigr)\ ,\hspace{.37cm} \eeqa
where $P(N,\eta)$ is the polynomial
\beqa P(N,\eta)= N^4+9 N^3+14 N^2+7
N+1 +4(N-1)\ \bigl(N^3+6N^2+5N+1\bigr)\,\eta\nn
+2\, \Bigl( N^2(N-1)\,(3 N+8)+ N^2+4 N+3\Bigr)\,\eta ^2
\nn+(N-1)\, \bigl(N(N-1)+1\bigr)\ \Bigl((N-1)\, \eta +4 (N+1)\Bigr)\, \eta ^3\,.\eeqa
When $u\to0$ it then follows from \eq{RR} that, $K\to \infty$ in cases (I), (II), (III) and (IV), and $K\to0$ in case (V). At the other endpoint,  $K\to0$ in cases (I), (II) and (IV), and
$z\to \infty$ in cases (III) and (V). Therefore,  cases (I), (II), (IV) and (V) have one singularity, and case (III) is a \st\ trapped between two singularities.

Note that at the singularity at $u=0$, $\rho,p\to0$ in cases (I), (II) and (III), and these singularities are of the kind that, because of matter vanishes at them,  we call { empty singularities}.

In contrast, at the other three singularities, i.e., $u=\infty$ in cases (II) and (V) and $u=0$ in case (IV), $\rho$ and $p$ diverge.

It is easy to see that the singularities at u = 0 are repulsive.  Indeed, if we consider  a test particle initially at rest near the singularity,  we get  from \eq{met3} that
\beq \left.\frac{d^2 u}{d\tau^2}\right|_{\tau=0}=-\frac{1}{2}\ g^{uu}\ \pd u \ln|g_{tt}|>0\eeq
if  $\eta>-\frac{3N+1}{N-1}$.

In our discussion, we have left aside the four values of $\eta$ that are at the endpoints of the regions, for the sake of readability  and completeness, the solutions for these cases are postponed to appendix \ref{B}.

It follows from \eq{eos}, \eq{presion} and \eq{alfa} that $p$  and  $\rho$ are proportional to $\alpha$, then we can get a vacuum solution from \eq{met3} if it limit $\alpha\to 0$ makes sense. But, we immediately see  that this  only occurs if also $\beta\to 0$ keeping their ratio fixed. Thus,  proceeding in this way and after the change of coordinates
\beqa \lab{met4}  1+\kappa z= u^{\frac{N+1}{\eta(N-1)+3N+1}}\ ,
\eeqa
we get from \eq{met3}  the vacuum solution
\beqa \lab{taub}    ds^2=-
\left(1+\kappa z\right)^{-2\left(\frac{N-1}{N+1}\right)}  dt^2 +  \left(1+\kappa z\right)^{\frac{4}{N+1}}\hspace{-1mm}\left((dx^{\mch
1})^2+\dots+(dx^{\mch{n}})^2 \right)+ dz^2,\nn
-\infty<t<\infty,\;-\infty<x^{\mch 1}<\infty,\;\dots,\;-\infty<x^{\mch N}<\infty,\;0< 1+\kappa z<\infty \ ,   \eeqa%
where $\kappa$ is an arbitrary constant. This is the generalization to higher dimensions \cite{L} of Taub's plane vacuum solution \cite{taub}. The new limit $\kappa\to0$, brings \eq{taub} to Minkowski spacetime.

Regarding the dust limit, it follows from  \eq{eos} that it corresponds to $\eta\to\infty$. In this limit, \eq{met3} becomes
\beqa  ds^2=- dt^2 +  (dx^{\mch 1})^2+\dots+(dx^{\mch{n}})^2 +\,\frac{\beta}{4\alpha}\ u^{-2}  \left(1-\beta u\right)^{-2}\, du^2 , \eeqa%
and  the change of coordinate $z=\sqrt{\frac{\beta}{4\alpha}}\ \ln\left(\frac{u}{1-\beta u}\right)$ shows that this is Minkowski spacetime. On the other hand, it follows from \eq{ro} that $p$ vanishes as $\eta^{-2}$ and  $\rho$ as $\eta^{-1}$ in this limit. Therefore, as expected, of course, there are not static dust solutions.

Summarizing, we have studied the main properties of the solutions with positive $\rho$ as a function of the parameter  $\eta$, these properties are listed in Table \ref{tabla}. In particular, we have found that empty repulsive singularities appear in the cases with $\eta >-1$. In a similar way, of course,  we may analyze the less interesting cases with negative $\rho$, but we are not going to do it here.

From now on, we restrict ourselves to the most physically interesting case I ($\eta>1$). We recall that, in this case, spacetime is semi-infinite and the range of the coordinate $u$  is  $0<u<1/\beta$.
On the other hand, \eq{RR} shows that there a singularity at $u=0$ and $K\rightarrow0$ as $u\rightarrow1/\beta$ ($z
\rightarrow\infty$) suggesting it is asymptotically flat at spatial infinite
in the $z$ direction.
It follows from \eq{ro} that pressure vanishes at $u=0$ and  $u=1/\beta$,  it is positive elsewhere and it has a maximum at
\beq
u=\frac{(\eta-1)(N-1)}{2\,\beta\,\bigl(\eta(N-1)+N+1\bigr)}\,.\eeq
Since $\rho$ and  $p$ vanish at the singularity,
we see that it is an  empty  singularity.
Thus,  the attraction of  matter curves \st\, in such a way that a repulsive
singularity arises in a  place free of matter.

In the following sections we shall study  movement of particles and propagation of waves in this spacetime.

\
\section{The geodesics }\lab{geo}

In this section we  study the geodesics in this \st, in the case  $\eta>1$. Since the components of the metric are
independent of $t$, $x^{\mch 1}$, \dots, and $x^{\mch N}$, the momentum covector components $p_t$, $p_{\mch 1}$, \dots, and $p_{\mch N}$ are constant along the geodesics. For timelike geodesics, we set $\mu^2=1$, ${\tilde{E}}=-p_t/m$ and $\tilde{p}_{\mch i}=p_{\mch i}/m$, and we choose $\tau$ to be the proper time; and for null ones, we set $\mu^2=0$, ${\tilde{E}}=-p_t$ and $\tilde{p}_{\mch i}=p_{\mch i}$, and we choose $\tau$ to be an affine parameter. So, we
can write \beqa
\frac{dt}{d\tau}&=&u^{\frac{2(N-1)}{\eta(N-1)+3N+1}}\,\left(1-\beta
u\right)^{\frac{2}{\eta-1}}\,
{\tilde{E}},\lab{dt}\\
\frac{dx^{\mch i}}{d\tau}&=&u^{-\frac{4}{\eta(N-1)+3N+1}}\,{\tilde{p}_{\mch i}}\,, \quad\quad 1\leq i\leq N,\\
\left(\frac{du}{d\tau}\right)^2&=&\frac{4\,\alpha}{\beta}\ \ u^{\frac{2\eta(N-1)+4N}{\eta(N-1)+3N+1}}\  \left(1-\beta
u\right)^{\frac{2 \eta}{\eta-1}}
\nn &\times&\left[u^{\frac{2}{\eta(N-1)+3N+1}}\,\left(1-\beta
u\right)^{\frac{2}{\eta-1}}\,{\tilde{E}}^2 -\mu^2-
u^{-\frac{4}{\eta(N-1)+3N+1}}\, \tilde{p}^2\right],\label{rebote}\, \eeqa
where $\tilde{p}^2=\tilde{p_{\mch 1}}^2+\dots +\tilde{p}_{\mch N}{}^2$.

The right hand side of (\ref{rebote}) cannot be negative, then it must hold that
\beqa \label{E2}{\tilde{E}}^2\geq\mathcal{V}(u):=
\frac{  u^{\frac{4}{\eta(N-1)+3N+1}}\,\mu^2\,+ \, \tilde{p}^2}    {{u^{ \frac{2(N+1)}{\eta(N-1)+3N+1 }} \left( 1 - \beta u \right) }^{\frac{2}{ \eta-1 }}}\ . \eeqa
Therefore, only vertical null geodesics touch the singularity at $u=0$ ($z
=z_0$) and reach $u=1/\beta$ ($z=\infty$). Whereas, as it is shown below, non-vertical null ones ($\tilde{p}^2>0$) as well as massive particles bounce
before getting to the singularity and bounce again before reaching $z=\infty$.

For the former case, the geodesic equation can be
integrated in  closed form. Indeed, when $\mu^2=0$ and
$\tilde{p}^2=0$,  we get from \eq{dt} and
 \eq{rebote}
\beqa \frac{dt}{du}&=\pm
&\sqrt{\frac{\beta}{4\,\alpha}}\  u^{\frac{3N-2}{\eta(N-1)+3N+1}-1}\,\left(1-\beta
u\right)^{-1}\,,  \eeqa
thus (see, for example, \cite{tablarusa}) we have
\beqa |t-t_0|&=& \sqrt{\frac{\beta}{4\,\alpha}}\
\frac{\eta(N-1)+3N+1}{3N-2}\ u^{\frac{3N-2}{\eta(N-1)+3N+1}} \nn &\times& _2F_1\!\left(\frac{3N-2}{\eta(N-1)+3N+1},1;1+\frac{3N-2}{\eta(N-1)+3N+1};\beta u\right)
. \eeqa

Note that
\beqa
|t-t_0|\propto \begin{cases}u^{\frac{3N-2}{\eta(N-1)+3N+1}}\rightarrow0&\text{as}\,\,\, u\rightarrow0\\
-\ln(1-\beta u
)\rightarrow\infty &\text{as}\,\,\, u\rightarrow1/\beta\,.
\end{cases}\eeqa
Therefore, a freely falling photon bounces off the singularity at time $t_0$, and then continues  its upward travel to $u=1/\beta$ ($z=\infty$).

It follows from (\ref{E2}) that the movement of non-vertical ($\tilde{p}^2>0$) photons or massive particles is constrained to the region where ${\tilde{E}}^2\geq\mathcal{V}(u)$. Since,  $\tilde{p}^2$ and $\mu^2$ are not both vanishing in these cases, the function $\mathcal{V}(u)$ is a positive continuous function of $u$ for $0<u<1/\beta$ and, since  $\eta>1$ and $N\geq2$, $\mathcal{V}(u)\rightarrow+\infty$ when $u\rightarrow0$ or  $u\rightarrow1/\beta$.

Moreover, $\mathcal{V}(u)$ turns out to be a convex function in this interval. It can be seen as follows. We can write $\mathcal{V}(u)$ as linear combination, with non negative coefficients, of two functions of the sort  $f(u)= u^{-a} ( 1-\beta u)^{-b}$ with $a,b>0$. Now, a straightforward computation yields
\beq f''(u)= \frac{u^{-a-2} ( 1-\beta u)^{-b-2}}{a+b}\left(ab+(1+a+b)\Bigl(a-(a+b)\beta u\Bigr)^2\right)\,,\eeq
which is clearly positive in $0<u<1/\beta$. Then,  $\mathcal{V}''(u)>0$ in this interval, and so there is one and only one point $u_0$ where  $\mathcal{V}(u)$ attains a positive minimum.

Note that, for massive particles, the value of $u_0$ depends on  $\tilde{p}^2$  besides  the parameters of the solution ($N$, $\eta$ and $\beta$); whereas for photons with $\tilde{p}^2>0$, $u_0$ is independent  of  $\tilde{p}^2$.

However, for any case and any value of ${\tilde{E}}^2>\mathcal{V}(u_0)$, we find two (and only two) turning points  $u_{min}$ and $u_{max}$ where ${\tilde{E}}^2=\mathcal{V}(u)$.

In the case of massive particles which are not moving horizontally ($\mu^2=1$ and $\tilde{p}^2=0$), we immediately see that $u=u_0$ is a stable equilibrium ``plane''. So a particle at rest in this place will remain at rest, and ${\tilde{E}}^2=\mathcal{V}(u_0)$.  On the other hand, if ${\tilde{E}}^2>\mathcal{V}(u_0)$ the particle bobs up and down between $u_{min}$ and $u_{max}$.

In the case of non-vertical geodesics ($\tilde{p}^2>0$),  if ${\tilde{E}}^2>\mathcal{V}(u_0)$  timelike geodesics as well as  null ones are wiggly worldlines where the coordinate $u$ oscillates between $u_{min}$ and $u_{max}$, while the horizontal movement is unbounded. Whereas, if ${\tilde{E}}^2=\mathcal{V}(u_0)$, the movement is horizontal and the geodesics are straight lines in the submanifold $u=u_0$. However, as mentioned above, the position of the equilibrium ``plane'' $u_0$ where horizontal geodesics occur, is independent of  $\tilde{p}$ for massless particles, but depends on it for massive ones. Nevertheless, it is not difficult to see, that the worldlines of horizontally-moving massive particles  approach to the massless ones in the hight energy limit ($\tilde{p}^2\gg 1$).

\section{Reflection of waves at the singularity}\lab{Cauchy}

After finding that all geodesics bounce off the singular boundary, let us turn our attention to propagation of waves. In particular, we shall explore the behavior of an incident wave pulse when it reaches the singularity.

To this end, in this spacetime, we consider the propagation of a
massless scalar field with Lagrangian density \beq\label{lag}
\lag= -\frac{1}{2}\nabla^a\varphi\, \cd a \varphi =
 -\frac{1}{2} g^{ab}\pd a \varphi\, \pd b \varphi\ . \eeq
 As usual, we obtain the field equations by requiring
that the action
\beq S=\int\lag(\cd a\varphi,\varphi,g_{ab})\sqrt{|g|}\ dt\,dz\,d^N\!\!x\ \eeq
 be stationary under arbitrary variations of the fields $\delta
\varphi$ in the interior of any compact region, but vanishing at its
boundary. Thus, we have
$$ \cd a\left(\dlag{ \cd a \varphi}\
\right)=\dlag{ \varphi}\ ,
$$
which, in our case, reads
\beqa\label{wave}
\cd a \nabla^a \varphi\ = \frac{\pd a \left(\sqrt{|g|}\,g^{ab}\pd b
\varphi \right)} {\sqrt{|g|}}=0. \eeqa
By using \eq{met3} we can explicitly write down the wave equation.
However, it suffices for our purposes  to consider the behavior of waves propagating in a neighborhood of the singularity. Thus, taking into account that near the singularity $\beta u\ll 1$, we can therefore neglect the terms  $\beta u$ in \eq{met3}. If, in addition, we make the change of coordinates $t\rightarrow C^{\frac{N-1}{2}}t$, $x^{\mch i}\rightarrow C^{-1}x^{\mch i}$ and $u\rightarrow z= C^{-N} u^{\frac{2N}{\eta(N-1)+3N+1}}$, where the constant $C$ is given by
\beq C= \left(\frac{16\,  N^2}{\beta \bigl(\eta(N-1)+3N+1\bigr)^2}\right)^{\frac 1{N+1}}\,, \eeq
we get that near the singularity the metric \eq{met3} behaves as
\beqa \label{met2} ds^2\approx- (\sqrt{\alpha}z)^{-1+\frac1N}\,
\left(dt^2-dz^2\right)+ (\sqrt{\alpha}z)^{\frac{2}{N}}\left((dx^{\mch
1})^2+\dots+(dx^{\mch{N}})^2 \right)\ .  \eeqa
The line element in the right hand side  corresponds to the vacuum solution \eq{taub}, this can be
readily seen through the change of coordinate $(\sqrt{\alpha}z)^{\frac{1}{N}}\to (1+\kappa z)^{\frac{2}{N+1}}$.

By using \eq{wave} and \eq{met2} we get the wave equation satisfied by $\varphi$ near the singularity
\beqa\label{wave2}
\pd
{tt}\varphi =\frac1z\, \pd z(z\, \pd z \varphi
)+\frac{1}{(\sqrt{\alpha}z)^{1+\frac1N}}\Delta_N \varphi\ , \eeqa
where $\Delta_N$ is the usual Laplacian  of the $N$-dimensional ``horizontal'' space.

The solutions of the wave equation \eq{wave2} have been studied in \cite{GST}. We  briefly describe here the main results.
We start with defining the underlying
elliptic differential operator $A$ and the Hilbert space $H$ on
which $A$ is symmetric.

Let $\Omega =\R^N\times (0,\infty)$ and $d\mu$ the Lebesgue measure on $\Omega $. Define, when $\varphi \in C_0^{\infty}(\Omega)$, the
operator $A$ by
\beq
A\varphi =-\frac{1}{z}\pd z\left(z\, \pd z \varphi
\right)-\frac{1}{(\sqrt{\alpha}z)^{1+\frac1N}}\Delta \varphi.
\eeq
Then consider the Hilbert space
\begin{eqnarray}
H&:=&L^2(\Omega, z\, d\mu)\nn &=&\{\varphi(x,z) \ :\ \int_\Omega
|\varphi(x,z)|^2\, z\, d\mu <\infty\}.\nonumber
\end{eqnarray}
By construction the operator $A$ is symmetric on $H$, with
\begin{eqnarray}
<A\varphi,  \eta>_H&=&\int_\Omega \left( \pd z \varphi \, \pd z
\bar{\eta} \,+ \frac{1}{(\sqrt{\alpha}z)^{1+\frac1N}}\nabla \varphi \cdot \nabla
\bar{\eta}\right)z\, d\mu\nn &=&:b(\varphi,\eta)\nonumber
\end{eqnarray}
 for $\varphi, \eta \in C_0^{\infty}(\Omega)$.

 This leads to introducing the ``energy space''
\begin{equation}\label{energyspace}
\mathcal{E}=\{\varphi \in H^1_{loc}(\Omega)\cap H \ :\,
b(\varphi,\varphi)<\infty\},
\end{equation}
where  $H^1_{loc}(\Omega)$ is the usual local Sobolev space. It is
straightforward to check that $\E$, equipped with its natural norm
 $$
 \|\varphi\|^2_{\E}:=b(\varphi, \varphi)+\|\varphi\|^2_{H},
$$ is a Hilbert space. This is the largest subspace of $H$ on which
the form $b$ is finite everywhere.

At the impossibility of provide a boundary condition at the singularity,
our first question is whether $A$ is essentially
self-adjoint or not: as a result, it is not. Indeed, we probed that it holds \cite{GST}\\
{\bf Theorem} {\em The operator $A$ is not essentially self-adjoint. However, there
exists only one self-adjoint extension of $A$ whose domain $D$ is
included in the energy space $\E$.}

However,
we are only looking for those extensions with domain included in
the energy space, because we are interested in waves having finite
energy. When taking into account this restriction, we recover the
uniqueness of the self-adjoint extension of $A$.
The domain of this particular
extension is
\beq
D:=\{\varphi \in \mathcal{E} \ : \exists\  C> 0\
\forall \ \eta \in \mathcal{E}, \ |b(\varphi,\eta)|\leq C \|\eta\|_H \}.
\eeq
Note that, there is no boundary condition attached to the definition of
$A$.

\medskip
Now, coming back to the wave equation, we take suitable functions
$f$ and  $g$ on $\Omega$ and consider the Cauchy problem

\beqa \lab{P}\left\{\begin{array}{ll}
 \pd {tt}\varphi(t,x^{\mch 1},\dots, x^{\mch N},z)=-A\varphi(t,x^{\mch 1},\dots, x^{\mch N},z),\\
\varphi(0,x^{\mch 1},\dots, x^{\mch N},z)=f(x^{\mch 1},\dots, x^{\mch N},z),\\
  \pd t  \varphi(0,x^{\mch 1},\dots, x^{\mch N},z)= g(x^{\mch 1},\dots, x^{\mch N},z).
\end{array}\right.
\eeqa

We found in \cite{GST} that, if $f\in \E$  and $g\in H$,  the problem \eq{P} has a unique solution,
and there exists a constant $C>0$ such that
\beq
\forall \ t>0\ \ \ \|\varphi(t,\cdot)\|_{\E}+ \|\pd t
\varphi(t,\cdot)\|_{H}\leq C(\|f \|_{\E}+ \|g\|_{H}).
\eeq
Furthermore, in this case, the energy
\beq
E(\varphi,t):=\frac{1}{2}\int_{\Omega} \left((\pd t
\varphi)^2+(\pd z \varphi)^2+\frac{1}{(\sqrt{\alpha}z)^{1+\frac1N}}\ |\nabla \varphi|^2
\right)z\,d\mu
\eeq
 is well-defined and conserved:
\beq
\forall\  t>0 \ \ \ E(\varphi, t)=\frac12 \left( \|g\|_H^2+ b(f,f)
\right).
\eeq

This result shows that the Cauchy problem \eq{P} is well posed
without any boundary condition on $\varphi$. This does not necessarily mean,
however, that $\varphi$ vanishes, or has no limit at all,
at the boundary. Indeed, provided $f$ and $g$ are regular enough,
$\varphi$ does have a trace on $\partial \Omega$ at each time $t>0$, which
is entirely determined by the Cauchy data (see \cite{GST}, for details).

In other words, if the energy of the initial field configuration is finite, it is conserved and, at any later time, the field is completely determined by the Cauchy data anywhere---even at the boundary. Therefore, no boundary condition should and can be provided.

It is important to remark, that this result may be rigorously obtained by using Theorem 7.1 of \cite{GST}, without making any approximation for $\beta u\ll 1$. If fact, by using \eq{met3} and \eq{wave} we can explicitly write down the exact wave equation. The relevant coefficient  $\sqrt{|g|}\,g^{uu}$, turns out to be $\sqrt{\frac{4\alpha}{\beta}}\,u (1-\beta u) $, and its inverse is not integrable at the origen.

Since the energy is conserved we see that waves are completely reflected at the empty singular boundary.

\section{Concluding remarks }

We have presented a  detailed analysis of the  solution of Einstein's equations for a static and hyperplane symmetric
perfect fluid satisfying the equation of state $\rho=\eta\, p$, for arbitrary $N\geq2$.

For $\eta>-1$, this exact solution clearly show how the attraction of distant matter
can shrink the \st\ in such a way that it finishes at an empty
singular boundary. Therefore, in spite of the weakening of gravity with the number
of dimensions of the space-time \cite{Tang,Cruz}, our solution
clearly shows that it is still strong enough to generate  empty
singular boundaries  for arbitrary $N\geq2$.

For $\eta>1$ we studied the properties of the geodesics, we found that only  vertical null geodesics touch the boundary and
bounce, and  all of them start and finish at $z=\infty$; whereas non-vertical
null  as well as all time-like ones are bounded
 between two planes determined by initial conditions.

We also  showed that the Cauchy problem for the propagation of a massless scalar field is well-posed, if we only demand the waves to have finite energy, although no boundary condition is required. And  waves are completely reflected at the singularity.

\begin{appendix}

\section{Killing vectors  and adapted coordinates}\label{A}

We want to find  coordinates adapted  to  a  hyperplane symmetric
distribution of matter. That is, \st\  must be
invariant under $n$ translations  and under  rotations in $n(N-1)/2$ hyperplanes.

More precisely, a $n+2$ dimensional \st\, will be said to be
$n$-dimensional Euclidean homogenous if it admits the $r=n(N+1)/2$
parameter group of isometries  $n$-dimensional Euclidean space
ISO(n).

Since the spacetime admits $n$  mutually conmuting independent
motions, we can choose coordinates $x^i,z,t$ so that the
corresponding Killing vectors are $\xi_{(i)}=\pd{i}$
($i=1,\dots,n$), and so \beq \xi_{(i)}^k =
\delta^k_i\ \ \ \ \ \ \
\left(\begin{array}{c}k=t,1,\dots,n,z\\i=1,\dots,n\end{array}\right)\,.\eeq
The equations of Killing, \beq \xi^k \pd {k}g_{ij}+ g_{kj}\ \pd
{i}\xi^k + g_{ik}\ \pd {j} \xi^k= 0\,, \label{killing}\eeq
corresponding to these vectors reduce to \beq \label{aa} \pd {k}
g_{ij}= 0\ \ \ \ \ \ \ \ \ \ \
\left(\begin{array}{c}i,j=t,1,\dots,n,z\\k=1,\dots,n\end{array}\right)\,.\eeq
Hence all the components of the metric tensor depend only on the
coordinates $t$  and $z$ and the metric is unaltered by the finite
transformation \beq x^i \rightarrow x^i + a^i\ \ \ \ \ \ \
(\text{for}\ i=1,\dots ,n)\,.\eeq We can take for the remaining
$n(N-1)/2$ motions the generators $\xi_{(ij)}= x^i \pd j - x^j \pd
i $ ($i<j\ \text{and}\ i,j=1,\dots,n$), so \beq \label{ab} \pd l
\xi_{(ij)}^k =\delta^i_l \delta^k_j-\delta^j_l \delta^k_i\ \ \ \ \
\ \  \left(\begin{array}{c}k,l=t,1,\dots,n,z\\i<j\ \text{and}\
i,j=1,\dots,n\end{array}\right)\,.\eeq Taking into account (\ref{aa})
and (\ref{ab}) from the equations of Killing (\ref{killing}) we
get \beq  g_{jm} \delta^i_l -g_{im} \delta^j_l+g_{lj}
\delta^i_m-g_{li} \delta^j_m= 0\ \ \ \ \ \ \
\left(\begin{array}{c}m,l=t,1,\dots,n,z\\i<j\ \text{and}\
i,j=1,\dots,n\end{array}\right)\,.\eeq From the last equation we
readily find \beq  g_{ii}=g_{jj} \ \ \ \text{and}\ \ \
g_{ij}=g_{it}=g_{iz}=0 \ \ \ \ \ \ \ (\ i\neq j\ \text{and}\
i,j=1,\dots,n)\,.\eeq Furthermore, we can take the bidimensional
metric of the $V_2$ spaces $x^i=$ constant ($i=1,\dots ,n$) in the
conformal flat form. Hence, the most general metric admitting this group of isometries  may be written as
 \beq  ds^2= - e^{2 U(z,
t)}\left(dt^2-dz^2\right)+ e^{2V(z, t)}\left((dx^{\mch
1})^2+\dots+(dx^{\mch{n}})^2
\right)\,.\eeq%

If, in addition we impose staticity, $U$ and $V$ must be time independent, and the change of variable $\int e^{U(z)}dz \rightarrow z$ brings  the line element to the form \eq{met}.

\section{Limit cases}\label{B}
In this appendix, we discuss the limit cases which were not considered section \ref{sol}.

\subsection{$\eta=1$}
By setting $\varepsilon=\eta-1$, $\alpha= \varepsilon a$ and $\beta=\varepsilon b$ in \eq{met3},
in the limit $\varepsilon\to0$, we readily get
\beqa  ds^2=-
e^{2b u}\, u^{-\frac{N-1}{2N}} \,dt^2
+u^{\frac{1}{N}} \left((dx^{\mch 1})^2+\dots+(dx^{\mch{N}})^2\right)
+\frac{b}{4a}\,  e^{2b u}\, u^{-\frac{3N-1}{2N}}\, du^2  \nn
-\infty<t<\infty,\;-\infty<x^{\mch 1}<\infty,\;\dots,\;-\infty<x^{\mch N}<\infty,\;0<u<\infty \,.\hspace{1.5cm}\eeqa%
Similarly, we get from \eq{ro}
\beqa  p(v)=\rho(v)=
\frac{(N-1)\ a}{2N\pi G}\  e^{-2 b u}\ u^{\frac{N-1}{2N}} \,,\eeqa%
and from \eq{RR}
\beqa K =\frac{a^2 \ e^{-4bu}}{{b^2\,N^2}\ u^{1+\frac{1}{N}}}\, \Bigl(N^2-1
+ 8N(N-1) \,b \, u +
16N(N+2) \,b^2 \, u^2\Bigr)\ .\eeqa
Then, since $a,b>0$ in this case, density and pressure vanish at $u=0$ and $u=\infty$, and the solution has an empty singularity at $u=0$. This solution is the higher-dimensional generalization of that of \cite{TT}.

\subsection{$\eta=0$}
It directly follows from \eq{met3} that, for $\eta=0$, depending on the sign of $\alpha$ and $\beta$, we get two very different solutions
\beqa   ds^2=-
u^{-\frac{2(N-1)}{3N+1}}\,\left(1-\beta
u\right)^{2}\, dt^2 + u^{\frac{4}{3N+1}}\left((dx^{\mch
1})^2+\dots+(dx^{\mch{N}})^2
\right) +\,\frac{\beta}{4\alpha}\ \ u^{-\frac{4N}{3N+1}}\ du^2,\nn
-\infty<t<\infty,\;-\infty<x^{\mch 1}<\infty,\;\dots,\;-\infty<x^{\mch N}<\infty,\;0<u\begin{cases}<1/\beta &\text{if}\ 0<\alpha,\beta<\infty\\ <\infty &\text{if}\ -\infty<\alpha,\beta<0\,.\end{cases} \hspace{.5cm}
\eeqa
In the first case, the solution has all the same  properties  as solutions in region (III) listed in Table \ref{tabla}. In the second case, it has all the same ones as those in (II).

The matter, in this case, is a weird fluid with null density but non-vanishing pressure, and  \eq{ro} becomes
\beq \rho(u)=0,\quad \quad  p(u)= -\frac{2\,(N-1)\,\alpha}{(3N+1)\pi G}\  \frac{u^{\frac{N-1}{3N+1}}}{\left(1-\beta u\right)}\,. \eeq

\subsection{$\eta=-1$}
In this case, \eq{met3} becomes
\beqa \lab{met4}  ds^2=-
u^{-\frac{N-1}{N+1}}\,\left(1-\beta
u\right)\, dt^2 + u^{\frac{2}{N+1}}\left((dx^{\mch
1})^2+\dots+(dx^{\mch{N}})^2
\right) +\,\frac{\beta}{4\alpha}\ \frac 1 {u \left(1-\beta
u\right)}\, du^2,\nn
-\infty<t<\infty,\;-\infty<x^{\mch 1}<\infty,\;\dots,\;-\infty<x^{\mch N}<\infty,\;0<u<1/\beta \,.\hspace{.5cm}
\eeqa
From \eq{ro} it follows
\beq \rho(u)=-p(u)= \frac{(N-1)\,\alpha}{2(N+1)\pi G}\,, \eeq
which shows that \eq{met4} turns out to be a vacuum solution with cosmological constant $\Lambda=\frac{2N}{(N+1)}\,\alpha$; this is the higher-dimensional generalization of that found in \cite{NH}. In this case the integral in \eq{z} can be computed in closed form, and we get that the depth of this spacetime, independently of $\beta$ and $N$, is $\frac{\pi}{2 \sqrt{\alpha}}$.

On the other hand, we get from \eq{RR}
\beqa K=\frac{16\ \alpha^2 \Bigr(N^2(N-1)+2(N+2)\, u^2\,\beta ^2\Bigl)}{{(N+1)^3\, u^2\,\beta ^2}}
\ , \eeqa
then, there is a singularity at $u=0$, and  $K=\frac{16\ \alpha^2 \Bigr(4+N(N-2)\Bigl)}{{(N+1)^2}}$ at $0=1/\beta$.

\subsection{$\eta=-\frac{3N+1}{N-1}$}
We now set $\varepsilon=\eta(N-1)+3N+1$, $\alpha= \varepsilon a$ and $\beta=\varepsilon b$,
and by making the coordinate change $v=1-\beta u$ and rescaling  $t$, $x^{\mch
i}$ and $\alpha$, we get
the limit $\varepsilon\to0$ of \eq{met3}
\beqa  ds^2=-
e^{2(N-1)\,b v}\, v^{\frac{N-1}{2N}} \,dt^2
+e^{-4b v} \left((dx^{\mch
1})^2+\dots+(dx^{\mch{N}})^2
\right)
+\frac{b}{4a}\,  e^{-2(N+1)\,b v}\, v^{-\frac{3N+1}{2N}}\, dv^2  \nn
-\infty<t<\infty,\;-\infty<x^{\mch 1}<\infty,\;\dots,\;-\infty<x^{\mch N}<\infty,\;0<v<\infty \,.\hspace{1.5cm}\eeqa%
Similarly,  \eq{ro} becomes
\beqa  p(v)=-\frac{3N+1}{N-1}\rho(v)=-
\frac{a(N-1)^2}{2N\pi G}\  e^{2(N+1)\,b v}\ v^{\frac{N+1}{2N}} \,,\eeqa%
and \eq{RR} gives
\beqa K =\frac{16}{N}\, a^2 \, e^{4(N+1)\,b v}\ v^{1+\frac{1}{N}}  \hspace{7cm}\nn \times\Bigl(2+N\bigl(13+N(9N-8)\bigr)
+ 8N^2(N-1)(3N+1) \,b  v +
16N^3(N^2-1) \,b^2  v^2\Bigr)\ . \eeqa
Therefore, since $a,b>0$, density and pressure vanish at $v=0$, but diverge at $v=\infty$; and the solution has a singularity at $v=\infty$.
\end{appendix}



\begin{thebibliography}{9}



\bibitem{q1}Gamboa Sarav\'{\i}, R.E.: Int.J.Mod.Phys. A {\bf 23}, 1995 (2008)

\bibitem{q2}Gamboa Sarav\'{\i}, R.E.: Class. Quantum Grav. {\bf 25}, 045005 (2008)

\bibitem{q3} Gamboa Sarav\'{\i}, R.E.:   Gen. Relativ. Gravit. {\bf 41}, 1459 (2009)

\bibitem{q4} Gamboa Sarav\'{\i}, R.E.: Int.J.Mod.Phys. A {\bf 24}, 5381 (2009)

\bibitem{GST} Gamboa Sarav\'{\i}, R.E., Sanmartino, M.,  Tchamitchian, P.:  Class. Quantum Grav. {\bf 27}, 215016 (2010)

\bibitem{Tang} Tangherlini, F.R.:  Nuovo Cimento {\bf27}, 636 (1963)

\bibitem{Cruz} Ponce de Leon, J.,  Cruz, N.:  Gen. Relativ. Gravit. {\bf32}, 1207 (2000)

\bibitem{Bron} Bronnikov, K.A.: J. Phys. A Math. Gen. {\bf 12}, 201 (1979)

\bibitem{BronKov} Bronnikov, K.A., Kovalchuk, M.A.: Gen. Relativ. Gravit. {\bf 11}, 343 (1979)

\bibitem{Coll} Collins, C.B.: J. Math. Phys. {\bf 26}, 2268 (1985)

\bibitem{SKMHH}
Stephani, H.,  Kramer, D., Maccallum, M.,  Hoenselaers, C., Herlt,
E.:  Exact Solutions to Einstein's Field Equations, Second
edn. Cambridge University  Press, Cambridge  (2003)

\bibitem{NH} Novotn\'y, J.,  Horsk\'y, J.:  Czech. J. Phys. B {\bf 24}, 718 (1974)


\bibitem{L} Liang, C.: J. Math. Phys. {\bf31}, 1464 (1990)


\bibitem{taub}
Taub, A.H.: Ann. Math. {\bf 53}, 472 (1951)


\bibitem{tablarusa}
Gradshteyn, I.S.,  Ryzhik, I.M.:  Table of Integrals,
Series, and Products, Academic Press, New York (1963)


\bibitem{TT}Tabenky, R.,   Taub, A.H.: Commun. Math. Phys. {\bf 29} 61, (1973)





\end{thebibliography}
\end{document}